\documentclass[iop,apj]{emulateapj}
\usepackage{amsmath,amssymb,amstext}
\usepackage[breaklinks,colorlinks,citecolor=blue,linkcolor=magenta]{hyperref} 

\usepackage[all]{hypcap}
\def\baselinestretch{0.96}
\usepackage{graphicx}
\usepackage{subfigure}
\usepackage{gensymb}

\begin{document}

\title{On the detection of Global 21-cm signal from Reionization using interferometers}

\author{Saurabh Singh$^{1,2}$}
\author{Ravi Subrahmanyan$^1$}
\author{N. Udaya Shankar$^1$}
\author{A. Raghunathan$^1$}
\affil{$^1$Raman Research Institute, C V Raman Avenue, Sadashivanagar, Bangalore 560080, India; saurabhs@rri.res.in}
\thanks{$^2$Joint Astronomy Program, Indian Institute of Science, Bangalore 560012, India }

\begin{abstract}
Detection of the global redshifted 21-cm signal is an excellent means of deciphering the physical processes during the Dark Ages and subsequent Epoch of Reionization (EoR). However, detection of this faint monopole is challenging due to high precision required in instrumental calibration and modeling of substantially brighter foregrounds and instrumental systematics. In particular, modeling of receiver noise with mK accuracy and its separation remains a formidable task in experiments aiming to detect the global signal using single-element spectral radiometers.   Interferometers do not respond to receiver noise; therefore, we explore here the theory of the response of interferometers to global signals.  In other words, we discuss the spatial coherence in the electric field arising from the monopole component of the 21-cm signal and methods for its detection using sensor arrays.   We proceed by first deriving the response to uniform sky of two-element interferometers made of unit dipole and resonant loop antennas, then extend the analysis to interferometers made of 1-D arrays and also consider 2-D aperture antennas. Finally, we describe methods by which the coherence might be enhanced so that the interferometer measurements yield improved sensitivity to the monopole component. We conclude that (a) it is indeed possible to measure the global 21-cm from EoR using interferometers, (b) a practically useful configuration is with omnidirectional antennas as the interferometer elements, and (c) that the spatial coherence may be enhanced and detectability of the global EoR signal may be smoothened using, for example, a space beam splitter between the interferometer elements.  
\end{abstract}


\section{Introduction}

Models for the cosmological thermal evolution in the baryons as a consequence of the first sources of radiation and reionization in our cosmic history are poorly constrained. Observational studies of the Epoch of Reionization (EoR) as well as the preceding Dark Ages are thus necessary to understand the formation of first stars and galaxies as well as the evolution of the diffuse intervening medium to its present state \cite[]{2001ApJ...563....1V, 2003MNRAS.343.1101C, 2006astro.ph..3149C, 2009RvMP...81.1405M}. There are various observational probes to study this epoch like the Gunn-Peterson effect, Cosmic Microwave Background, quasars, Gamma ray bursts etc. \cite[]{2006ARA&A..44..415F}. However, most are limited in value due to their being integral measurements or because they involve relatively difficult NearInfrared (NIR) observations \cite[]{2013PhDT.......168P}.  The measurement of the global or all-sky  redshifted 21-cm from the spin flip transition of HI perhaps represents the most direct probe of baryons during the Dark Ages and subsequent EoR making it the ``richest of all cosmological data sets" \cite[]{2005ApJ...624L..65B}.

There have been several theoretical studies that model these epochs and derive predictions for the nature of the redshifted 21-cm global signal \cite[]{2006PhR...433..181F, 2008PhRvD..78j3511P} and also suggest the astrophysics that might be derived by its measurement \cite[]{2010PhRvD..82b3006P,2013ApJ...777..118M,2014Natur.506..197F}. There are many ongoing experiments that attempt to detect the global 21-cm signal using single antenna elements: EDGES \cite[]{2010Natur.468..796B,2008ApJ...676....1B}, SARAS \cite[]{2013ExA....36..319P}, LEDA \cite[]{2015ApJ...799...90B}, SCI-HI \cite[]{2014ApJ...782L...9V} and BIGHORNS \cite[]{2015PASA...32....4S}.  However, the detection of this signal remains unsuccessful to date because the design of a spectral radiometer with the required accuracy in calibration of systematics is a formidable challenge.  Additionally, the recovery of the EoR global signal, which has maximum amplitude less than 100~mK, requires specialized methods to distinguish it from Galactic and extragalactic foregrounds of several 100~K. 

Motivated by the formidable challenge of discriminating against instrument related internal systematics in single-element radiometers, there has been recent work on interferometer based detection of the global signal \cite[]{2014arXiv1407.4244V, 2014arXiv1406.2585M,2015arXiv150101633P}.  Compared to single-element radiometers, interferometers are relatively insensitive to receiver noise and noise originating internally in ohmic losses and passive components in the signal path.  The work presented herein develops the theory of the response of interferometers to the global 21-cm signal and explores a variety of configurations that may usefully make interferometer measurements of the global spectrum.  The configurations include measurements of the spatial coherence in the electromagnetic field owing to the global signal as well as methods that enhance this coherence so as to improve the detection sensitivity.

Recent studies have also shown that ionospheric refraction and absorption may add excess power which could be 2--3 orders of magnitude greater than the signal of interest \cite[]{2014MNRAS.437.1056V, 2014arXiv1409.0513D}.  This consideration is a compelling argument for observations to be made from above the atmosphere and from space where the response is free of ionospheric distortions; therefore, the configurations we consider here are assumed to be in space.  Nevertheless, the conclusions arrived at here following the analyses and comparisons apply equally well for ground based interferometers.

\break
\section{Notations and preliminaries}
\label{not}
We begin by clarifying the notations used throughout this paper.  We consider interferometer measurements of the global 21-cm and hence the interferometers and methods considered herein operate at radio frequencies.  In all cases, we consider here the response of \textit{two-element interferometers}; therefore, any reference to interferometers refers to two-element interferometers only.  Any two-element interferometer measures the spatial and temporal coherence between the fields at two spatially separated locations at which sensors are positioned.  The pair of sensors in a two-element interferometer are called the \textit{elements of the interferometer}; the interferometer elements are \textit{antennas}. The term \textit{baseline} refers to the relative spacing and orientation of the interferometer elements; baseline is a vector.  

The antenna, which is the interferometer element, may in practice be a single sensing unit such as a dipole antenna or resonant loop: we refer to such antennas as \textit{unit} antennas.  The antenna may be a 1-D phased array of such units.  The antenna may be 2-D phased array of units, or a 2-D aperture made of reflectors along with sensors at the focus that act together as concentrators of the electromagnetic (EM) field. 

The antennas essentially sense the EM field at their location and provide a weighted summation of the EM field over the antenna area or \textit{aperture}; a voltage waveform corresponding to the net field is provided at the antenna terminals and the two-element interferometer measures the coherence between such voltage waveforms sensed by a pair of elements.  We use the term \textit{response} to refer to the response of an interferometer to the global signal unless stated otherwise.   It may be noted here that the effective aperture of an antenna might be larger than the physical aperture.

Finally, although the detection method discussed here is relevant to the monopole component of any astronomical signal, our signal of interest is specifically the all-sky or uniform component of the redshifted 21~cm from HI in the Epoch of Reionization, which is referred to as the 21-cm monopole or the global 21-cm signal. 

While considering this uniform component, we assume a sky across which the emission is uniform but spatially incoherent.  For such a sky, the square of the voltage at the antenna terminals represents the average brightness temperature over the beam power pattern or radiation pattern of the antenna, which represents the relative sensitivity of the antenna over sky temperature.  As an illustrative example, we may consider an antenna whose planar aperture is a collection of unit dipoles that are combined in an impedance matched network to yield the net voltage at the antenna terminals. In this case, all the dipoles would sense the same rms voltage at their spatial locations owing to the uniform sky, and the output would have the same rms voltage as the rms voltages sensed by the individual dipoles.  This is required by thermodynamics considerations. The output power has fractional contributions from all parts of the aperture; the output power is a weighted average of the aperture powers, where the weighting is by the aperture illumination.  In summary, for a sky across which the emission is uniform and incoherent, the antenna has an aperture that defines an area over which the antenna does a weighted averaging of the field strength to provide a voltage at its terminals.

For a uniform sky that is incoherent across angle on the sky plane, we may define the spatial coherence function in the visibility domain to be the mutual coherence in fields sensed or sampled by antennas with isotropic beam patterns.  The response of an interferometer made of such isotropic antennas is what we define to be a `true' coherence.  This `true' coherence function has a value at the origin of the visibility plane that is the brightness of the uniform sky.  Assuming identical antenna elements, the interferometer response is an integral of the coherence function over a visibility-plane footprint of a shape that is the auto-correlation of the element aperture.  This footprint is centered at the location of the baseline vector on the visibility plane.  

If the baseline length is less than the effective diameters of the apertures, then the footprint will cover the origin and hence the integral response would include a substantial response to the brightness of the uniform sky.  Otherwise, the integral will always be less than the sky brightness, and might be expected to be smaller for longer baselines and larger aperture sizes if not zero.  
 
\section{Response of a two element interferometer to a global signal}
\label{2el}

Interferometers measure the spatial coherence function \citep[]{1999ASPC..180....1C} of the electromagnetic field.   And it is commonly believed that interferometers are sensitive only to brightness temperature variations on the sky and do not respond to the uniform or monopole component.  Therefore, interferometers and Fourier synthesis telescope arrays are usually used in astronomy to measure the spatial coherence owing to discrete sources of radiation on the sky, and thereby indirectly image the  source structures and brightness variations.     

In contrast, here we focus instead on the  spatial coherence that is due to the monopole component of the sky brightness distribution.  We present a study of the  expected variation in the coherence with changing baseline as well as with observing frequency.  While it is indeed true that by and large interferometers are `blind' to the uniform sky, we show below that there are special circumstances in which interferometers might usefully respond to the monopole component of the sky brightness distribution.

The response $V(\vec{b}, \nu)$ of an interferometer to sky brightness distribution $T_{sky}(\vec{r},\nu)$ is a function of the baseline vector $\vec{b}$ and frequency $\nu$ (or equivalently the wavelength $\lambda$) \citep[]{thompson2008interferometry}:   
\begin{equation}
V(\vec{b}, \nu)=\frac{1}{4\pi}\int_{\Omega} A(\vec{r},\nu) T_{sky}(\vec{r},\nu)e^{-i 2\pi \frac{\vec{b} \cdot \vec{r}}{\lambda}}d\Omega.
\end{equation}
The integral here is over the entire sky, with $\vec{r}$ representing position unit vector towards solid angle element $d\Omega$ on the sky.  $A(\vec{r},\nu)$ represents the beam power pattern of the interferometer elements.  It is assumed that the interferometer elements constituting the 2-element interferometer are identical.

For a signal that is global in nature and uniform over the sky, $T_{sky}(\vec{r},\nu)$ may be written as just $T_{sky}(\nu)$ and taken out of the above integral, which may then be written as
\begin{equation}
V(\vec{b}, \nu)=\frac{1}{4\pi} T_{sky}(\nu) \int A(\vec{r},\nu) e^{-i 2\pi \frac{\vec{b} \cdot \vec{r}}{\lambda}}d\Omega.
\label{eq:visint}
\end{equation}

If $T_{sky}(\nu)$ is in units of Kelvin, then the response $V(\vec{b}, \nu)$ is also in Kelvin units.  As shown below, this integral is nonzero. Indeed, for short-spacing interferometers the integral may be a substantial part of the mean brightness temperature of the sky, which indicates that interferometers may be configured to have a substantial and useful response to the global redshifted 21-cm signal. We compute this integral below for different types of interferometer elements.

\subsection{Interferometers made of unit antennas}
\label{sec:element}

In this subsection, we compute Equation~\ref{eq:visint} for four cases in which the interferometer elements are unit antennas. 

\begin{figure}[ht]
\centering
\subfigure[Parallel configuration ]
{
\includegraphics[scale=0.5]{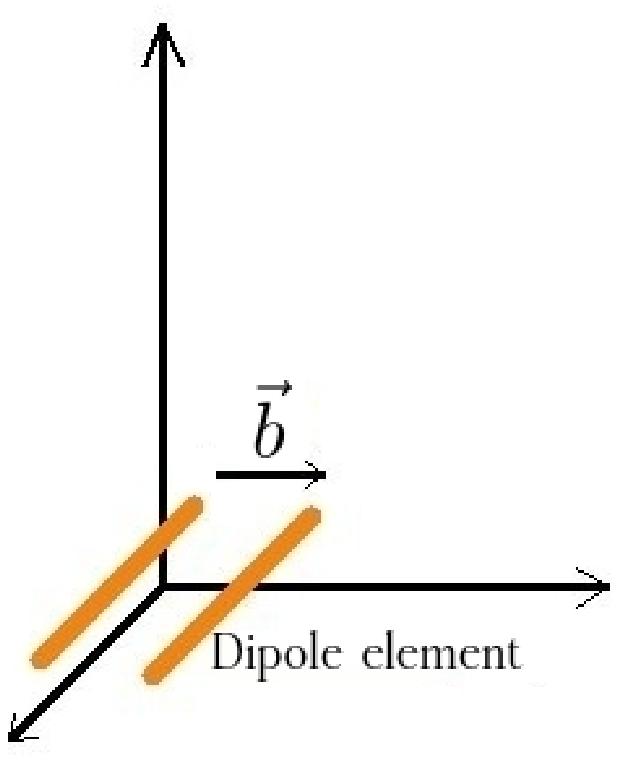}
\label{fig:s2}
}
\quad
\subfigure[In-line configuration]
{
\includegraphics[scale=0.5]{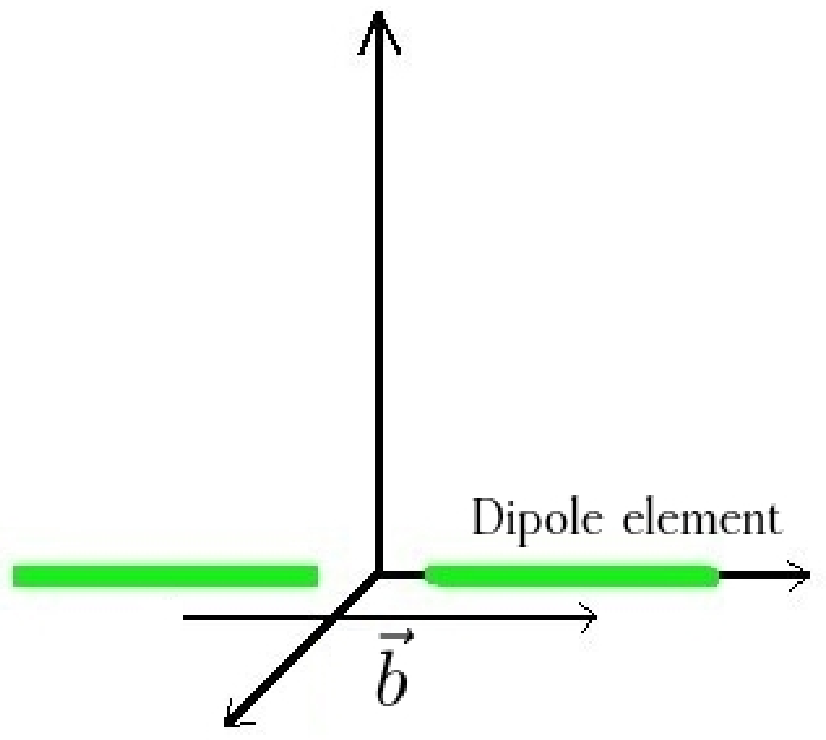}
\label{fig:s1}
}
\caption{Configuration for two element interferometers.}
\label{fig:geom}
\end{figure}

In the first two cases, the interferometer elements are assumed to be identical short dipoles at the observing frequency, with lengths much less than $\lambda/2$, where $\lambda$ is the wavelength of the observation. The radiation pattern of the short dipole is of toroidal form with nulls along the axis of the dipole, with response of the form sin$^2(\theta$), where $\theta$ is the angle measured from the axis. In the first case, the axes of the pair of antennas are oriented to be parallel to each other and perpendicular to the baseline vector, as depicted in the figure in Panel~(a) of Fig.~\ref{fig:geom}.  In the second case the interferometer elements are once again assumed to be identical short dipoles but with their axes oriented along the baseline vector; this configuration is depicted in Panel~(b) of Fig.~\ref{fig:geom}. We call these first and second cases as `parallel' and `in-line' configurations respectively.  

In the third case the elements are assumed to be circularly-polarized resonant loop antennas tuned to the observing frequency, with the loop axes orthogonal to the baseline vector.  The circumference of the loops is equal to the observing wavelength and the antenna patterns for the resonant loops are of cos$^2(\theta$) form, where $\theta$ in this case is the angle from the axis of the loop antenna. 

For reference, we also compute Equation~\ref{eq:visint} for the case where the interferometer elements are isotropic antennas.

\begin{figure}[h]
\begin{center}
\includegraphics[scale=0.35]{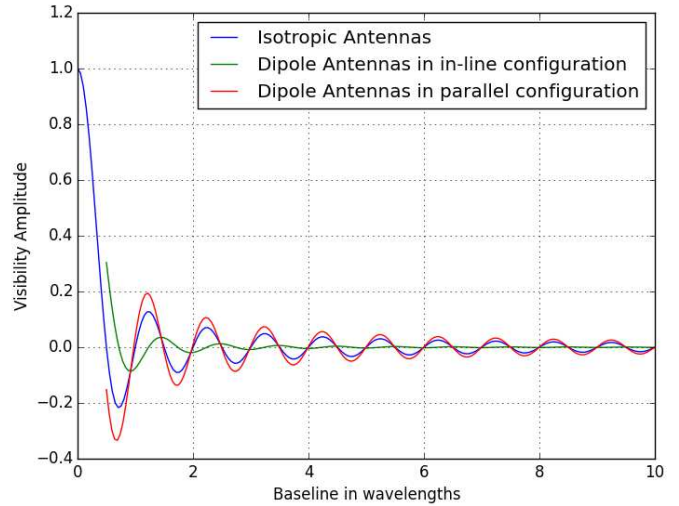}
\caption{Response to uniform sky of a 2-element interferometer made of identical unit antennas. The response of a resonant loop antenna is identical to the case of short dipoles in in-line configuration and their traces overlap. In-line and parallel configuration responses have been traced from $\rm{0.5}\lambda$ avoiding the near field regions of the antennas.}
\label{fig:Visibility1}
\end{center}
\end{figure}

We show in Fig.~\ref{fig:Visibility1} the response of the interferometer versus baseline length for these four cases. All plots are normalized to the value at a baseline length of zero, which is the value that a conventional total-power measurement using a single antenna element would yield for a uniform sky.  Isotropic antennas or antennas with isotropic radiation patterns are not realizable in practice, they notionally correspond to point sensors of the field.  As discussed earlier, the trace in Fig.~\ref{fig:Visibility1} corresponding to isotropic antennas represents a `true' spatial coherence in the field arising from a uniform sky brightness. 

First, there is substantial response of the interferometers to uniform sky - interferometers can indeed measure a global signal.   At zero length baseline, this coherence represents the autocorrelation or power in the field from the uniform sky.  With increasing baseline length the spatial coherence in the field falls off substantially; in fact, the spatial coherence is a sizable fraction of the total power only for separations less than a wavelength.  This is consistent with what is known in optics of the coherence properties of the radiation field in a cavity filled with blackbody radiation \citep[]{1964PhRv..134.1143M}.

The response in the case of dipoles in parallel configuration is greater than that for the isotropic case, and the response for in-line dipoles is smaller than for isotropic; the response in the case of resonant loop antennas is same as that for dipoles in in-line configuration.  As seen in Fig.~\ref{fig:Visibility1}, for baselines of a few wavelengths, the peak response in the case of parallel dipoles is about a factor of five greater than that for in-line dipoles.  However, the response amplitude is strongly dependent on the baseline length, fluctuating about zero and reducing with increasing baseline length as in a damped sinusoid, and the amplitude and the amount of damping of the amplitude with increasing baseline length are both strongly dependent on the nature of the interferometer elements.

Since the coherence in the field varies fairly rapidly with the baseline, varying by close to a period for a change in baseline length of a wavelength, integrating over visibility domains comparable to or greater than a wavelength would substantially diminish the net  interferometer response.  This decrease in the response would be more pronounced if the aperture has a greater extent along the baseline vector, since it is in this direction that the coherence in the field varies.  Dipoles in in-line configuration have a greater effective extent along the baseline vector compared to dipoles in parallel configuration; it is for this reason that the interferometer response of two-element interferometers with dipoles in in-line configuration have relatively lower response.  

Response to uniform sky is a maximum when the baseline length is zero.  An alternate physical understanding for the cause of the interferometer response to uniform sky may be arrived at by examining the effective area afforded in directions where the projected baseline is zero.  Dipoles in parallel configuration have maxima along this zero-baseline direction and nulls in the orthogonal direction towards which the projected baseline is a maximum.  Short-dipole interferometers in in-line configuration, as well as interferometers with elements that are resonant loops, have nulls in their beam patterns along the baseline vector in the direction where the projected baseline is zero; therefore it is unsurprising that these configurations have a smaller response to uniform sky compared to the case of the parallel configuration.

It may be noted here that we have assumed that the interferometers are in space, with no ground.  If the interferometer is placed above ground, and the ground below the antennas are covered with ideal absorbers, the sky response of the interferometer and that of the total-power of a single antenna would both be halved,  without any change in the normalized visibility functions.

\subsection{Interferometers made of 1-D antenna arrays}
\label{sec:1Darray}

We next extend the analysis to interferometers whose elements are 1-D linear arrays consisting of short dipoles.  The short dipoles that form the units of the 1-D antenna are assumed to be arrayed along the length of the antenna; i.e., their linear polarizations are aligned to be along the length of the 1-D antenna.  We also assume that the signals from the units of the 1-D antennas are combined with zero phase difference and equal weights to provide the voltage signal at the terminals of the antennas.  Because the dipole units are collinear and arrayed along the length of the antenna, and because antennas with such a configuration have isotropic radiation patterns in the plane perpendicular to the axis along which the units are arrayed, we refer to such interferometer elements as 1-D antennas. 

We consider a linear array of $N$ identical dipole units spaced $d = (\lambda/2)$ apart.    As stated above, in the plane perpendicular to the antenna axis, the 1-D antennas have omnidirectional radiation patterns.  In any plane containing the axis, the net far-field radiation pattern is obtained by multiplying the radiation pattern of a single unit with an Array Factor:
\begin{equation}
\label{eq:array}
AF=\frac{1}{N}\left[\frac{{\rm sin}(\frac{N\psi}{2})}{{\rm sin}(\frac{\psi}{2})}\right].
\end{equation}
Here $\psi = (2 \pi / \lambda)~ d~{\rm cos}(\theta)$, where $\theta$ in this case is the angle from the long axis of the 1-D array \citep[]{Balanis:2005:ATA:1208379}. The Array Factor is maximum along directions perpendicular to the 1-D array.  

\begin{figure}[ht]
\centering
\subfigure[Parallel configuration]
{
\includegraphics[scale=0.35]{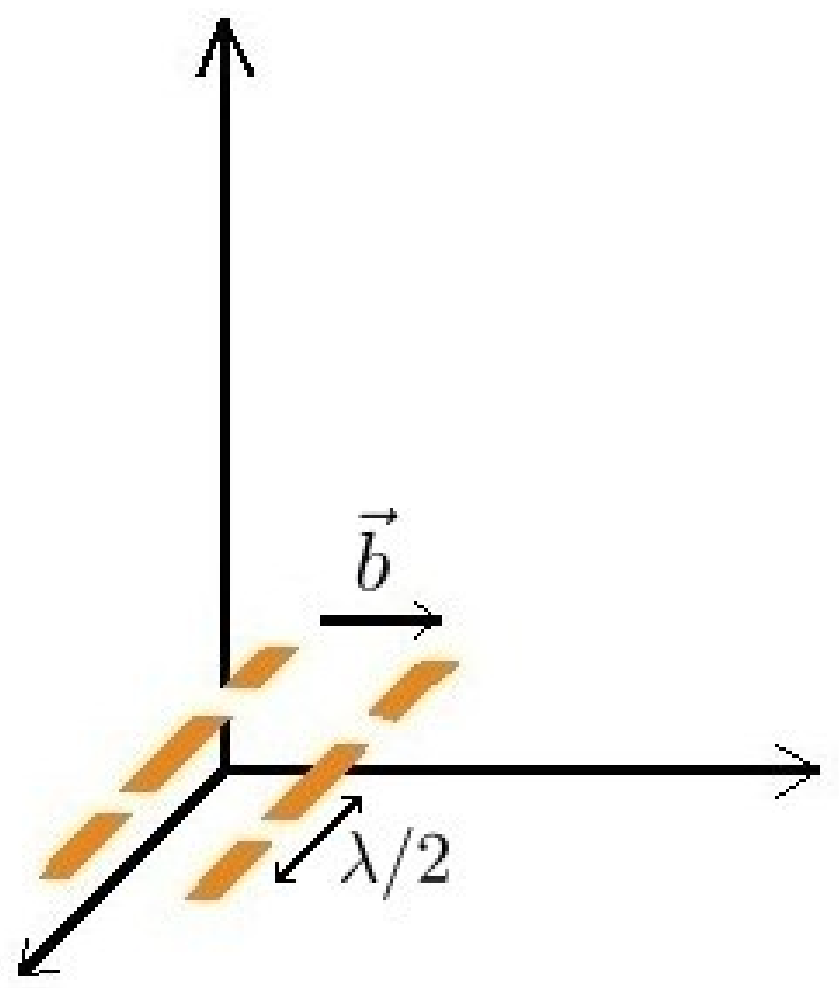}
\label{fig:s22}
}
\quad
\subfigure[In-line configuration ]
{
\includegraphics[scale=0.46]{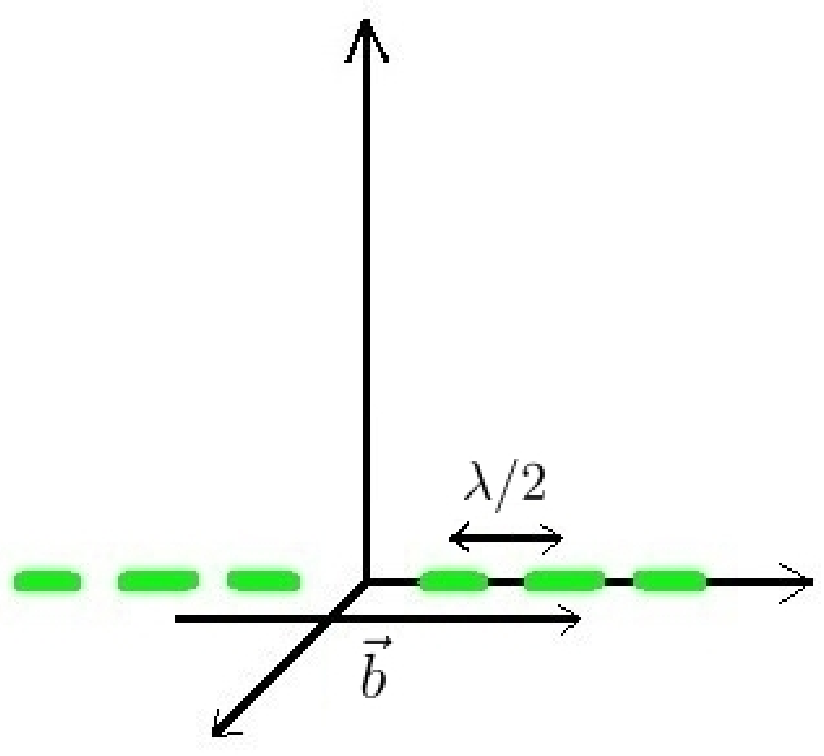}
\label{fig:s11}
}
\caption{Configurations for two-element interferometers consisting of 1-D arrays as interferometer elements.}
\label{fig:geom2}
\end{figure}

We consider two cases in this category: one in which the 1-D antennas are perpendicular to the baseline vector, a parallel configuration, and a second case in which the 1-D antennas are along the baseline vector, which is an in-line configuration. The geometries for both cases are shown in Fig.~\ref{fig:geom2}.

\begin{figure}[htbp]
\begin{center}
\includegraphics[scale=0.45]{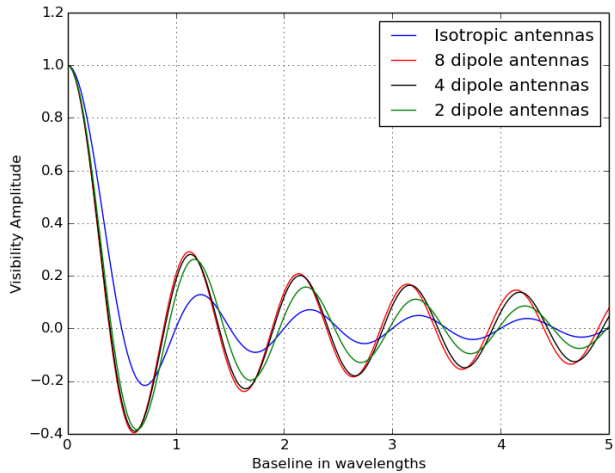}
\caption{Interferometer response in the case of 1-D antennas in parallel configuration, for antennas with different numbers of dipoles.}
\label{fig:i_p}
\end{center}
\end{figure}

\begin{figure}[htbp]
\begin{center}
\includegraphics[scale=0.33]{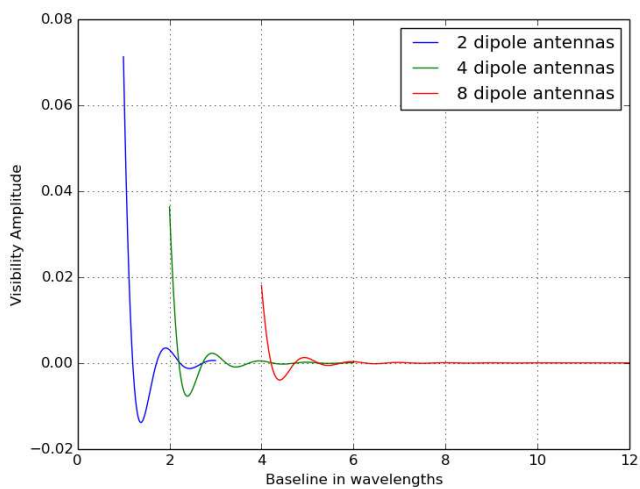}
\caption{Interferometer visibility amplitude versus baseline length for antennas with in-line arrays in in-line configuration (for antennas with different numbers of dipoles). The visibilities are normalized to give the fractional response to the global sky brightness temperature.}
\label{fig:uniform}
\end{center}
\end{figure}

In each of these two cases we compute the response to uniform sky as a function of baseline length and for different numbers of short dipole units within the 1-D antennas.  Fig.~\ref{fig:i_p} shows the response of the parallel configuration versus baseline length, in this figure the response to isotropic antennas is also shown for reference. The corresponding plot for the in-line configuration of 1-D antennas is in Fig.~\ref{fig:uniform}. 

First, the in-line configuration does not admit close packing and small baselines because of overlap and shadowing.  Therefore, the shortest baseline in the case of the in-line configuration of 1-D antennas is equal to the length of the 1-D antenna elements, which is larger when the antennas are made of greater numbers of units.  When the shortest baseline is larger, the maximum response, which occurs when the baseline is smallest, is diminished.  For this reason, in-line configurations are inherently poorer in sensitivity compared to parallel configurations.

The limiting baseline is either set by geometry, as discussed above, or the size of reactive zones of the interferometer elements. If a pair of antennas were placed close to each other and within their respective reactive zones, they would suffer significant mutual coupling. For any antenna of dimension $D$, operating at wavelength $\lambda$, the reactive zone is considered to be within a radial distance of $\frac{D^2}{\lambda}$, and baselines are best maintained to well exceed this size if the individual antenna performances are to be unperturbed by proximity to their neighbor.  In the case of the parallel configuration the system performance is better defined when the interferometer elements are separated by more than their reactive zones, which sets the minimum baseline. 

The visibility amplitude in the case of the parallel configuration is greater than that for the case of isotropic antenna elements, where as the response of the interferometer with in-line configuration is relatively small and also diminishes more rapidly with increasing baseline length.   As in the case for unit dipole antennas as elements of the interferometer, this is consistent with the expectation that averaging of the baseline-dependent complex coherence over longer baseline lengths results in diminishing of the response.

The response falls rapidly with increasing number of units in the case of 1-D interferometer elements in an in-line configuration.  The result may be understood by arguments similar to those presented in Section~\ref{sec:element}.  Adding more units in an in-line configuration directly increases the extent of the aperture in the radial direction in the visibility plane along which the complex coherence varies most rapidly.  Additionally, the domain of the integration is over a one-sided radial segment of the complex coherence function that does not include the origin.  Therefore, any increase in the extents of the 1-D antennas beyond about half a wavelength results in a substantial diminishing of the integral response.  In the alternate perspective discussed above, increasing the numbers of units in the in-line antennas increases the gains of the interferometer elements, narrows the beam pattern to be more directed in the plane perpendicular to the axis, which results in reduced response towards the direction in which the projected baseline is zero.  

On the other hand, increasing the number of short dipole units within the 1-D antennas in the parallel configuration tends to increase the sensitivity of the array to the global signal.   Increasing the number of units in this case extends the 1-D array, and hence the integration over the coherence function, in a direction tangential to the baseline vector.  Most importantly this integral is over a domain that is two sided in which the coherence function is symmetric.   Therefore, for small increases in numbers of units the response is enhanced; however, as the numbers of units grows and the length of the 1-D antennas is substantially greater than the baseline length the integral yields diminishing returns in terms of increased response.  In the alternate perspective, increasing the numbers of units in the 1-D antennas oriented perpendicular to the baseline increases the gain towards the direction where the projected baseline is zero, reducing the response in orthogonal directions, and this may be viewed as causing the enhanced response to uniform sky.

\subsection{The case of aperture antennas}

We next consider interferometers between antennas with circular apertures.   This case has been discussed previously by \citet[]{2015arXiv150101633P} and we comment on their analysis below at the end of this Section.  In this case study the antennas may be 2-D aperture arrays or reflectors with focal feeds.  We describe the aperture antennas using a function $g(u)$ that describes the field distribution on the aperture plane.  We assume circular symmetry in this field distribution and that the field $g(u)$ may be expressed as a function of the distance $u$ from the center point only.  Therefore, the far field radiation pattern of the aperture antenna may be computed as a Radial Fourier Transform, also known as Hankel Transform, of the aperture field distribution:
\begin{equation}
\label{eq:fourier}
F(\theta)=2\pi\int_{0}^{u_{max}} u J_0(2\pi u \sin\theta)g(u)du.
\end{equation}
Here $u$ is expressed in wavelengths and $u_{max}$ is the radius of the circular aperture in wavelengths.  $F(\theta)$ is the far-field voltage radiation pattern; $\theta$ here is the offset angle in radians from the axis of the aperture.  $J_{0}$ is the Bessel function of zeroth order.   

We consider aperture antennas of two descriptions: one in which the sensor of the field provides a uniformly weighted summation over the aperture plane and a second in which the field in the aperture is added with an amplitude weighting corresponding to a Gaussian taper. Since the aperture is of finite size, even for the case where the aperture field is averaged with a Gaussian taper the far-field radiation pattern cannot be of Gaussian form; instead, the pattern would be the Fourier Transform of a truncated Gaussian.

\begin{figure}[htbp]
\begin{center}
\includegraphics[scale=0.33]{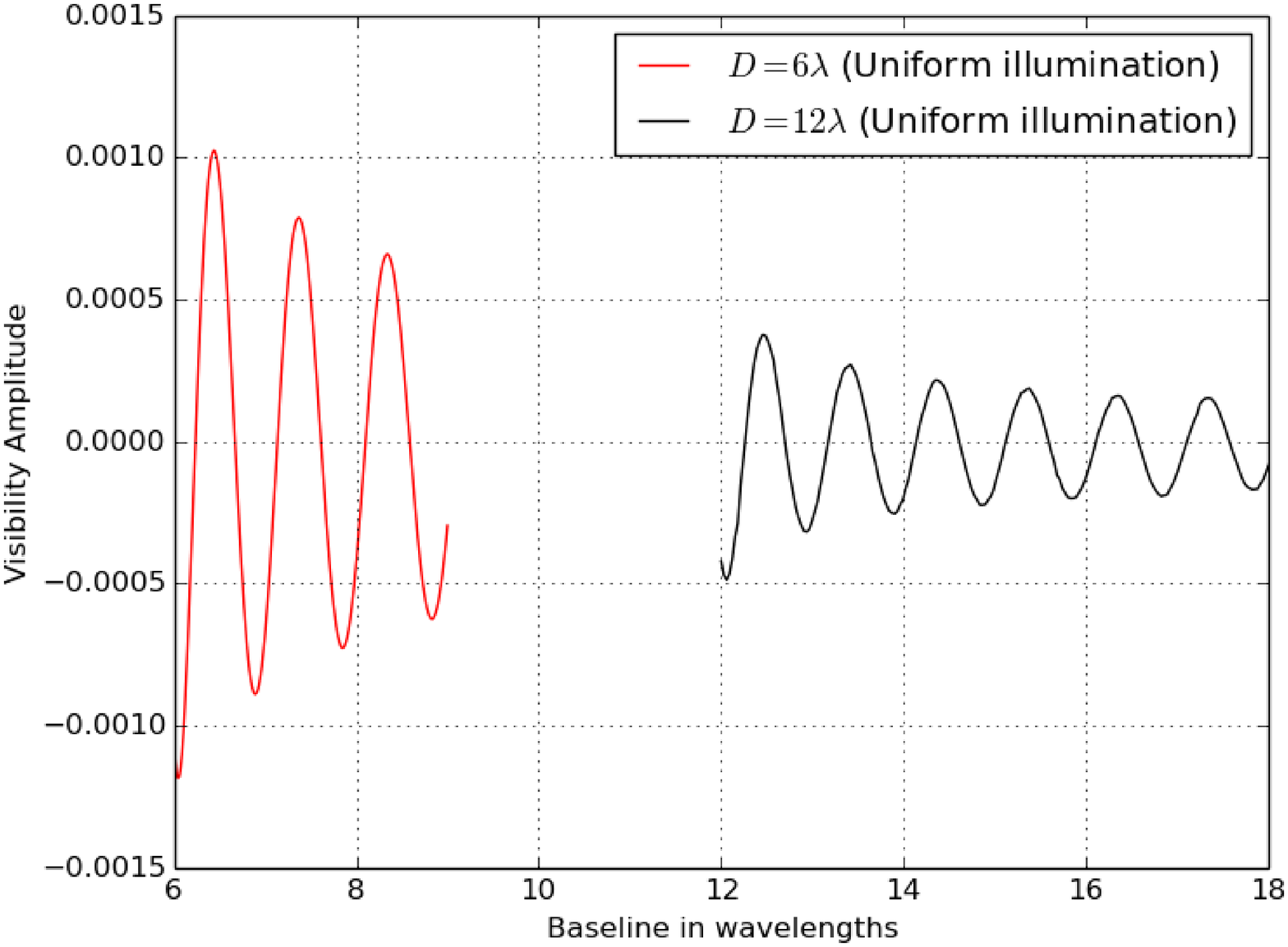}
\caption{Interferometer visibility amplitude versus baseline length for circular aperture antennas that have a uniform sampling of their aperture fields.}
\label{fig:ap}
\end{center}
\end{figure}

Using $F(\theta)$ from Equation~\ref{eq:fourier} as the response function of the antenna elements, 
we may now use Equation~\ref{eq:visint} to compute the response to a global sky brightness for an interferometer made from a pair of circular apertures.  In Fig.~\ref{fig:ap} we show this response for the case of uniform weighting of the field over the antenna aperture.  We show the responses for the cases where the aperture diameters $D$ are $6 \lambda$ and $12 \lambda$.  The response is only shown where the baseline exceeds the aperture diameter since smaller baselines are impossible without overlap and hence shadowing.  The magnitude of response to global sky is at most about $10^{-3}$ of the global sky brightness; additionally, the visibility amplitude diminishes with increasing dish size and increasing baseline length. 

For antenna apertures of diameter $D$ as the elements of an interferometer, the integration of the coherence function is over regions of diameter $2D$ in the visibility plane.  As discussed earlier, any integration over a region of the visibility plane that exceeds half a wavelength in size would substantially diminish the response of such an interferometer to the global 21-cm signal because (a) the coherence of the signal varies substantially with baseline length and (b) the footprint on the interferometer response on the visibility plane does not include the origin.  Aperture antennas with diameters exceeding a few wavelengths would have little response to the global mean brightness of the sky because they provide such spatially integrated measures of the coherence function.  This averaging over the varying complex coherence function, over domains that are substantially offset from the origin, is the cause for the substantial reduction in response in the case of aperture antennas.

Most often a tapering is used to down-weight the fields at the edges of the aperture while averaging to provide the voltages at the terminals of conventional aperture antennas.   This is done so that the antenna beam patterns have lower sidelobes and hence unwanted off-axis response is reduced.  In our second case study of two-element interferometers with aperture antennas we assume Gaussian form tapers of the aperture fields, in which the field at the aperture edges are down weighted to 10\% of the central value.   We find that the interferometer response to global mean sky  is furthermore reduced in this case relative to the uniform weighting case. For apertures of diameter $6 \lambda$, the visibility amplitude is below $10^{-7}$ at about the closest baseline length of $6 \lambda$, and diminishes further with increasing aperture size and baseline length.  This may be understood as because in any short-spacing interferometer formed between aperture antennas, the coherence function is a maximum at the shortest spacings that occur between the edge portions of the two apertures that lie closest, and any kind of edge tapering of the aperture fields would result in this contribution to the averaging being most down weighted.    In summary, interferometers made using 2-D aperture antennas are clearly substantially less sensitive to the global EoR signal compared to interferometers using 1-D antennas or unit antennas.

It has been pointed out earlier in \citet[]{2015arXiv150101633P} that the EoR monopole signal resides at the origin of the visibility plane of interferometers, and what is required is for an interferometer response to be sensitive to the origin.  It is also suggested therein that a primary beam of aperture elements could cause the response to sample this origin and, therefore, make an interferometer sensitive to the monopole.  As discussed above, the visibility-plane footprint of an interferometer has the size and shape of the autocorrelation of the antenna aperture; therefore, to get the origin into the visibility-plane footprint of an interferometer would require an antenna diameter $d$ exceeding the baseline length. To achieve this with aperture antennas, the two antennas forming the interferometer would have to overlap or shadow. No interferometer made of finite aperture antennas, which do not overlap or shadow, could possibly sample the origin of the visibility plane.  The primary beam profile assumed in \citet[]{2015arXiv150101633P} has been argued to be realistic and the response function in the visibility plane,  as computed from the adopted beam pattern, has been shown to sample the origin.    
This is only possible if the effective apertures of the antennas are larger than the physical apertures and the sampling of the origin of the visibility plane arises from overlap of the effective apertures.
Our view is that interferometers with finite aperture antennas do respond to the uniform sky, not because they sample the origin of the visibility plane, but because the coherence function corresponding to a uniform sky does extend away from the origin and may be sampled by aperture antenna interferometers.

\section{Enhancement of the spatial coherence corresponding to a uniform sky }
\label{sec:enhance}

\citet[]{2015MNRAS.450.2291V} suggested using lunar occultation of the uniform sky to generate and enhance the spatial coherence corresponding to a global sky signal, which may then be detected using interferometers.  In so far as the global redshifted 21-cm signal is concerned, blocking the sky with the Moon creates a disk shaped source, with diameter equal to the lunar disk, with a relative brightness temperature equal to the difference between the brightness of the lunar disc and the brightness of the global redshifted 21-cm signal.  The spatial coherence in the field corresponding to this differential disk source is what is proposed by \citet[]{2015MNRAS.450.2291V} to be detected using interferometers.  In this section, we discuss another technique to enhance the spatial coherence and hence the response of interferometers to any global signal.

\begin{figure}[htbp]
\begin{center}
\includegraphics[scale=0.35]{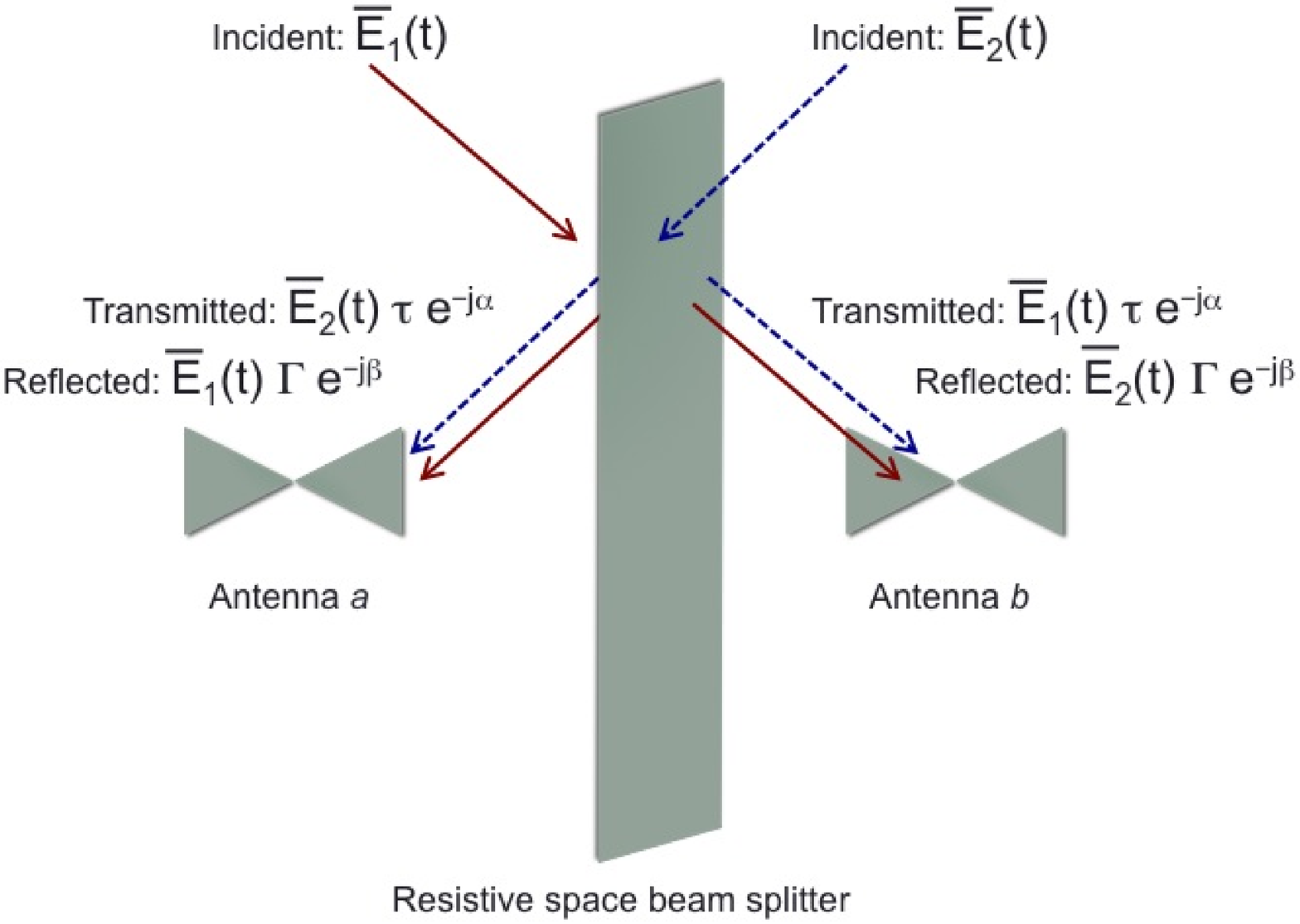}
\caption{Schematic of a configuration with a beam splitter sheet in between the interferometer elements \citep[]{2014arXiv1406.2585M}.}
\label{fig:zeb}
\end{center}
\end{figure} 

Any beam splitter that partially reflects and partially transmits incident electromagnetic radiation results in fields on the two sides that have a mutual coherence, which may be measured using an interferometer whose elements are placed on the two sides of the beam splitter.  We show in Fig.~\ref{fig:zeb} a configuration in which a space beam splitter is placed in between antenna elements of a two-element interferometer: the pair of antennas receives sky radiation that is partially transmitted through the sheet from the far side and partially reflected off the sheet from the near side.   Sky radiation is incident on the two sides from any uniform component of the sky and the reflected and transmitted fields that are sensed by the antenna elements now have a substantial mutual coherence.  This coherence would be well above that without a beam splitter in between.  The performance of space beam splitters was analyzed in \citet[]{2014arXiv1406.2585M} where it was shown that the sheet impedance was required to be resistive and of value half the impedance of free space (377/2~$\ohm$) for maximum coherence and hence interferometer response.  \citet[]{2014arXiv1406.2585M} also proposed a method for the construction of such a screen as a resistor grid, and demonstrated consistency between measurements of its performance with expectations based on electromagnetic modeling.

In a space beam splitter, the enhancement of spatial coherence in the fields corresponding to global signals may be alternately understood as follows.  As viewed from any sky direction the antenna element on the far side is seen through the screen and a reflected image of the antenna on the near side is seen to be coincident with the former.  In effect, the interferometer elements appear from all directions on the sky to present a zero length baseline.  This sampling of the origin of the visibility space may be considered, in this case where a beam splitter sheet is placed between the antenna elements, to be the cause of the enhanced response to global components of sky brightness.

\section{The sensitivity of small interferometer arrays to wideband global signals}

We consider below the spectral sensitivity of interferometers,  based on useful configurations emerging from the above discussions, to measure the global EoR signal over the 40--200~MHz frequency range.  The signal is assumed to be of 10~mK amplitude and the telescope system temperature is assumed to be dominated by the antenna temperature $T_a$, which is the sky brightness temperature modeled as a function of frequency $f$ as:
\begin{equation}
T_a=200\left(\frac{f}{150~{\rm MHz}}\right)^{-2.5}~{\rm K}.
\label{eq:ta}
\end{equation}

In the above discussions we have considered responses as function of baseline length; however, here we use those results to infer the response as function of frequency for interferometers that have fixed baselines.   A single baseline would have a frequency response--- the telescope response or `\textit{telescope filter function}'---that would have substantial variation over the 1:5 band, including null response at some frequencies.  Adding baselines of different lengths would avoid nulls in the net response.  We have chosen, as an undemanding illustration, to consider a very small array of three interferometer elements, indeed the smallest possible.  The first two are assumed to be spaced $\lambda_{max}$ apart and the third is at a distance of $1.5\lambda_{max}$ from the second, where $\lambda_{max}$ is the longest wavelength of interest, corresponding to 40~MHz. This configuration gives three baselines of length $\lambda_{max}$, $1.5 \lambda_{max}$ and $2.5 \lambda_{max}$.   This distribution of spacings ensures that visibilities are sampled at $(b/\lambda)>1$, where $b$ is the baseline length, at all frequencies. Thus mutual coupling, which is most severe when adjacent interferometer elements are within the reactive near fields of neighboring elements, is reduced.  The spacings between the interferometer elements is a trade off between deleterious mutual coupling and desirable signal power, both of which are greater at shorter baselines.  

The analysis in Section~\ref{sec:enhance} suggests that amongst the different antennas that might be elements of an interferometer, a 1-D antenna oriented perpendicular to the baseline vector, i.e. an in-line array in parallel configuration, has a better response to global sky signals. Hence we first consider below 1-D antennas made as an array of short wideband dipoles in parallel configuration (as shown in Fig.~\ref{fig:s22}), then consider 1-D antennas that are designed and constructed to be wideband 1-D apertures fully filled over the operating frequency range. Finally we consider the broadband response of a two-element interferometer with a space beam splitter in between two dipoles (as discussed in Section~\ref{sec:enhance}); we consider only the case of an in-line interferometer (as shown in Fig.~\ref{fig:s1}) since this configuration would have minimum mutual coupling and cross talk, which result in spurious unwanted responses. We refer to this last configuration as a zero-spacing interferometer.

\subsection{Very small interferometer array of 1-D antennas made of short dipoles in parallel configuration}
\label{sec:1D}

The antennas in this interferometer configuration are assumed to be linear arrays of collinear short dipoles spaced half wavelength apart at 40~MHz, so that the spacing in wavelengths would only be greater at all other frequencies in the band of interest. As discussed in Section~\ref{sec:1Darray}, since the improvement in gain diminishes substantially with increasing number of short dipoles in the 1-D antenna, we fix the number of dipoles to be four in each antenna of the interferometers.

We now estimate the effective signal-to-noise ratio (SNR) as a function of frequency.  Let $m_i$ denote the measurement set recorded in the $i^{\rm th}$ interferometer baseline and $r_i$ denote the telescope filter function or interferometer response for that baseline. An estimate of the global sky signal is given by $({m_i}/{r_i})$.  We then compute a weighted average of the estimates made in the different baselines, optimally weighting the estimates by the inverse of the noise variance, which is proportional to $r_i^2$.  This weighted average estimate of the signal $X_{\rm eor}$ is given by:
\begin{equation}
\label{eq:weight}
X_{\rm eor}=\frac{\sum\limits_{i=1}^3 m_i r_i}{\sum\limits_{i=1}^3 r_{i}^2},
\end{equation}
where the summations are over corresponding frequency data in the three baselines.

In any frequency channel, the rms noise uncertainty in the weighted mean estimate $X_{\rm eor}$ of the global EoR signal is given by
\begin{equation}
\label{eq:vari}
\sigma_{\rm eff}=\sqrt{\sum\limits_{i=1}^3 \frac{\sigma_{\rm noise} r_{i}}{\sum\limits_{i=1}^3 r_i^2}},
\end{equation}
where $\sigma_{\rm noise}$ is the rms noise in that channel.  We assume here that $\sigma_{\rm noise}$ is the same in all baselines and is dominated by the antenna temperature $T_a$ corresponding to the foreground brightness temperature (Equation~\ref{eq:ta}).  $\sigma_{\rm noise}$ is given by \cite[]{Wilson:52076} $\sigma_{\rm noise}^2 = \frac{T_{\rm b}^2}{2\beta\tau}$. We have assumed a channel bandwidth $\beta$ of 1~MHz and integration time $\tau$ of 200~hr.

The ratio of weighted mean estimate $X_{\rm eor}$ of the global EoR signal (Equation~\ref{eq:weight}) and the effective rms noise $\sigma_{\rm eff}$ (Equation~\ref{eq:vari}) yields the effective SNR for the telescope.

\subsection{Very small interferometer array made of 1-D aperture antennas}
\label{sec:1D aperture}

The 1-D antennas in  Section~\ref{sec:1D}  were linear arrays of short dipoles, spaced half wavelength apart at 40~MHz.  At this frequency the linear antenna is a fully filled 1-D aperture; however, at higher frequencies in the 40-200~MHz band the filling is increasingly sparse.  In this section we consider, as the interferometer elements, 1-D aperture antennas that are fully filled at all frequencies in the band.  This is indeed practically realizable by arraying small and wideband sensor elements all along the 1-D aperture so that the fields may be coherently combined with uniform weighting.  The 1-D aperture antennas are assumed to form interferometers in parallel configuration.

The effective SNR versus frequency is shown in Fig.~\ref{fig:snr} for the 3-element interferometer telescope.  Separate lines show the SNR for the case where the 1-D antenna is an array of dipoles spaced half wavelength apart at 40~MHz and the case where the antenna is a 1-D aperture.  Unsurprisingly, the 1-D aperture antenna improves upon the sensitivity at the higher frequencies (see Fig.~\ref{fig:snr}).

\subsection{Zero Spacing Interferometer}
\label{sec:zebra}

\begin{figure}[htbp]
\begin{center}
\includegraphics[scale=0.45]{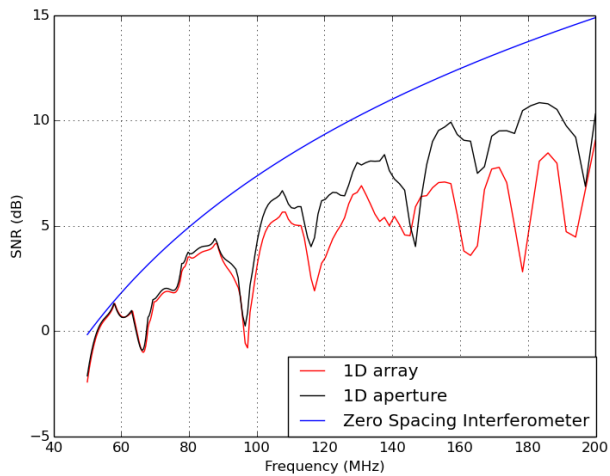}
\caption{Effective signal-to-noise ratio for the detection of a global signal of amplitude 10~mK.  The interferometer array is assumed to consist of three interferometer elements with three baselines formed between the elements;  the configuration of the in-line interferometers and 1-D elements are as described in the text. Also shown is the signal-to-noise ratio for a zero-spacing interferometer: a 2-element in-line interferometer of unit dipoles with a resistive sheet in between.  200~hr integration time and 1~MHz  spectral bandwidth are assumed.}
\label{fig:snr}
\end{center}
\end{figure}

We finally consider the wideband response of a zero spacing interferometer.  The interferometer elements in this case consists of short wideband dipoles and the interferometer is of in-line configuration.   A resistive sheet is in between the in-line dipoles and serves as a space beam splitter.  As discussed in \citet{2014arXiv1406.2585M}, for a resistive sheet with sheet impedance equal to half the impedance of free space, the reflected and transmitted powers received by an interferometer element are equal and each is one-fourth of the incident power.  Further, half the incident power is absorbed in the resistive sheet.  Assuming that the resistive sheet is sufficiently large in extent and the antennas are wideband, the interferometer response is frequency independent and the telescope filter function is a constant at 0.25. The SNR for such a zero-spacing interferometer is also shown in Fig.~\ref{fig:snr}.

\section{Discussion and Summary}

First, it is clear that all-sky spectral signals that are uniform across the sky, like the global EoR signal that is otherwise known as the EoR monopole, is detectable using interferometer methods, which have their inherent advantages over single element total power spectral radiometers.  2-element interferometers made of unit dipole elements or 1-D antennas that are composed of an array of short wideband dipoles do capture up to about 20\% of the global signal on baselines of a few wavelengths. 

Second, owing to the extremely small response to the global EoR signal of interferometers made using aperture antennas, any attempt at interferometer detection of global EoR ought to be done with elemental or 1-D antennas.  The response of interferometers made of small aperture antennas, with diameters 6--12 $\lambda$, and with uniform weighting in their sensing of the aperture fields, have a response that is less than $10^{-3}$ of the global EoR.  If the element apertures have a realistic Gaussian taper in their sensing over their apertures, then this response drops to lower than $10^{-7}$.  Since the system noise in interferometers at the frequencies at which the global EoR signal appears is dominated by the sky foreground brightness, interferometers made using aperture antennas would require at least 10$^4$ times greater observing time making them unattractive in comparison.

The spatial coherence in the field arising from the global EoR signal may be enhanced using a semi-transparent screen.  The response of any two-element interferometer to global EoR may be enhanced by placing a resistive screen in between, with sheet resistance equal to half the impedance of free space (377/2~$\ohm$).  The interferometer then senses the altered fields on the two sides of the screen, whose coherence has been enhanced by the screen.  The elements of the interferometer may now be a pair of short wideband dipoles oriented in in-line configuration, so that their mutual coupling and hence cross talk is minimized.  A critical advantage of global EoR measurements using such a zero-spacing interferometer is that its telescope filter function is relatively smooth compared to the net function derived from a small array of unit or 1-D antennas.  

It may be noted here that interferometers also respond to angular structure in sky brightness distribution and this response depends on the spatial frequency mode corresponding to the baseline length.  Since this is frequency dependent, interferometers mode-couple angular structure in brightness distribution to frequency structure in the spectral domain.  This results in confusion to the global EoR signal.  Placing interferometers EW, and averaging the response over time, removes the spectral structure arising from this mode coupling.

Antenna elements that have frequency dependent radiation patterns also mode couple angular structure in brightness distribution to the spectral domain. Therefore, it is advantageous to use only frequency independent antennas as interferometer elements.  This is yet another argument against using 2-D aperture antennas.  This is also an argument against using 1-D aperture antennas, and hence the antennas may simply be electrically-short wideband dipoles. 

The work presented here advances the understanding of the usefulness of interferometers in measurements of global EoR.  The work motivates in depth study of issues related to mutual coupling in short spacing interferometers and the consequent systematics and limitations to sensitivity.  Additionally, careful modeling of the response of interferometers with finite-size resistive sheets in between is suggested as future work, including the response to emission from the resistive screen itself.

\end{document}